# Analysis of Twilight Background Polarization Directions across the Sky as a Tool for Multiple Scattering Separation

Oleg S. Ugolnikov[1], Igor A. Maslov[1,2]

[1]Space Research Institute, Russian Academy of Sciences, Russia
[2]Moscow State University, Sternberg Astronomical Institute, Russia

Corresponding author e-mail: ougolnikov@gmail.com

The direction angle of twilight sky background polarization in the celestial sphere points far from solar vertical is found to depend on the ratio of single and multiple scattering contributions. The polarization direction behavior during the twilight is related with general properties of background components and can be used to check the accuracy of their separation method. The basic assumptions of this method are confirmed by analysis made in this paper. This helps to hold the temperature and scattering medium study in the upper mesosphere, the least accessible layer of the Earth's atmosphere. The mesosphere temperature data obtained during the observations from 2011 till 2014 are also presented.

## 1. Introduction

Mesosphere remains the worst investigated layer of the Earth's atmosphere. Physical conditions make it hard for *in situ* measurements, and the basic information is obtained by the remote sensing methods from the ground or from space. This study is of great importance since the mesosphere is also the region of rapid changes currently observed. The negative temperature trend is possibly the fastest over the whole Earth [1], and the summer temperature minimum reached in polar and mid-latitudes makes these atmospheric regions the coldest on Earth. Upper mesosphere is also the place of meteor particles moderation that leads to formation of metal and dust layers. Small dust particles are playing the role of condensation nuclei for polar mesospheric or noctilucent clouds those appear during the summer cold epoch and observed in Europe since 1885.

Non-equilibrium thermal conditions of the mesosphere require the different methods of temperature monitoring to retrieve the complete thermal structure of this layer. The basic data are provided by satellite measurements, two missions are currently operating: TIMED/SABER [2] and EOS Aura/MLS [3]. Ground-based methods (lidar, radar, optical and microwave sounding, etc.) give the additional data on the local mesosphere. This paper is devoted to the least expensive method of local measurements – the twilight sounding. The possibilities of the twilight study were described as early as in [4] and [5], but the results were affected by the sufficient contribution of multiple scattering of the solar emission in the lower atmosphere during the twilight period, especially its dark stage when the single scattering layer crosses the mesosphere.

Numerical simulation of multiple scattering [6] shows good agreement of results with observations data but remains model-dependent and can't be hold during dark period of twilight, when the effective order of scattering rapidly increases. But multiple scattering can be separated on the experimental basis, using frequent polarization measurements in different sky points, those can be done by all-sky CCD-camera with polarization filter correctly installed in the optical scheme. The basic ideas of this method are presented in [7], and its work range covers the stage of twilight with solar zenith angles from 97° to 100° and corresponding altitudes from 70 to 85 km – the region inaccessible for numerical analysis. In [8] the method was improved and used to build the temperature profiles of the upper summer mesosphere. The accuracy of single measurement was about 5K and the results were in good agreement with satellite data [2, 3].

General method description is performed in the same paper [8]. For the procession, it is necessary to point out the twilight stage when single scattering vanishes and the whole background consists only of multiple scattering. Basing on the dependencies of sky polarization during the twilight, it was found that this threshold is being crossed at the solar zenith angle equal to 99° in the zenith. This value linearly increases to 100° in the dusk area (sky point zenith angle 50° towards the Sun) and decreases to 98° in the opposite direction (at the same sky point zenith angle). It corresponds to the effective illuminated twilight layer baseline altitude about 90 km, that can be considered as the higher limit of the twilight analysis. This is the basic free parameter of the method. Its value needs additional confirmation, that will be done below.

In paper [7] only the solar vertical points were considered. In the following work [8] the method was expanded to the most observable part of the sky, adding the points with single scattering angles closer to 90° and high degree of single scattering polarization. However, these side sky points are also characterized by different polarization directions of single and multiple scattering – the phenomena vanishing in the solar vertical and not used in separation method. In this work we will analyze this event, its relation with the observable properties of the twilight sky and possible use for the updated separation procedure.

**2. Observations**

This paper is based on Wide-Angle Polarization Camera (WAPC [7, 8]) observations started in central Russia (55.2°N, 37.5°E) in summer 2011. General device description is made in [8]. The measured values are intensity, value and angle of polarization of different sky points with zenith angles up to 70° from the day till the deep night. The measurements are hold in a spectral band with effective wavelength equal to 540 nm.

Figure 1 describes the coordinate system used during the data procession. The sky point is characterized by angles $\zeta$ and $\tau$, where the module of last one is equal to the angular distance between this point and the solar vertical (unlike [8], it is positive westwards from the Sun).

Actually, this is a polar coordinate system where poles are placed at the horizon in 90° from the Sun. It is convenient since all points with equal $\zeta$ (away from poles) are characterized by the same dependencies of effective single scattering altitude on solar zenith angle [8]. Moreover, they have practically the same ratio of single and multiple scattering. The sky background intensity and polarization are found on 5°-grid by $\zeta$ and $\tau$.

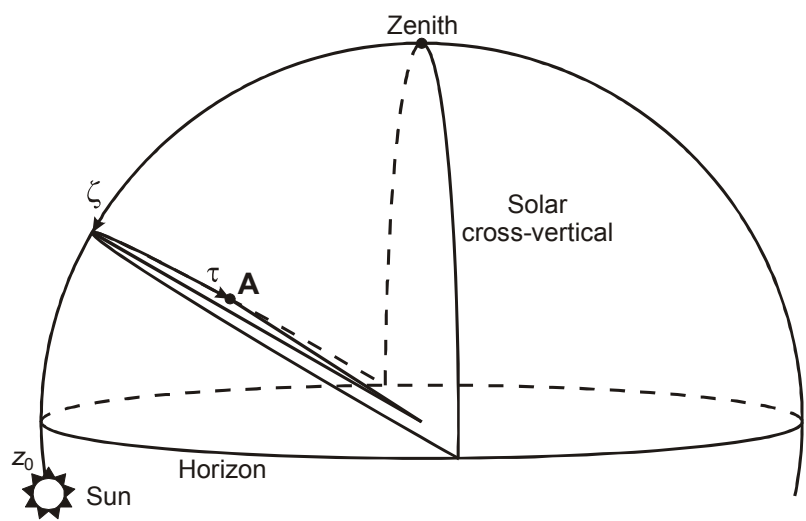

*Figure 1. Sky coordinate system used during the observations and data procession.*

The observational sky properties will be reviewed on the example of moonless evening twilight of June, 5th, 2013, with excellent weather conditions, high polarization value during the light twilight stage (about 0.77 in the zenith). The single scattering analysis by the method [8] shows that it was described by Rayleigh law with the accuracy better than 1% in altitude range 70-80 km for this twilight.

**3. Solar vertical and cross-vertical sky points polarization properties.**

Following [8], we denote the sky background polarization as $Q$ (not to mix it up with pressure value used there) and start from the dependency of $Q$ on the solar zenith angle, $z_0$, for a number of sky points in solar cross-vertical ($\zeta=0$), including the zenith ($\zeta=\tau=0$), that is shown in Figure 2. The values are close to each other, showing almost the same single and multiple scattering intensity ratio. The polarization rises a little away from the zenith (larger module of $\tau$) that can seem surprising. This effect can not be explained by the difference of single scattering angles, since this difference is too small, and the angles themselves are close to 90°. During the transitive twilight ($z_0>95°$) when increasing contribution of multiple scattering leads to polarization fall, this $Q$ difference become more significant prior to the maximum during the dark stage of twilight ($z_0>99°$) when single scattering totally disappears. Obviously, this polarization difference is the property of multiple scattering. It can be qualitatively explained using Figure 3.

Since the sky brightness increases from the zenith to the horizon, the secondary light source for multiple scattering can be considered as a thick horizontal ring in the sky (however, the sky brightness along the ring is not constant). For multiple scattering in the zenith, different parts of the ring will correspond to different polarization directions, and the total polarization decreases (it would even vanish if the ring had a constant brightness). Moving down along the solar cross-vertical, we see that light scattering from most part of the ring (including the points at the distance 90° giving the maximum contribution to the polarization) leads to the vertical polarization direction, almost the same as for single scattering.

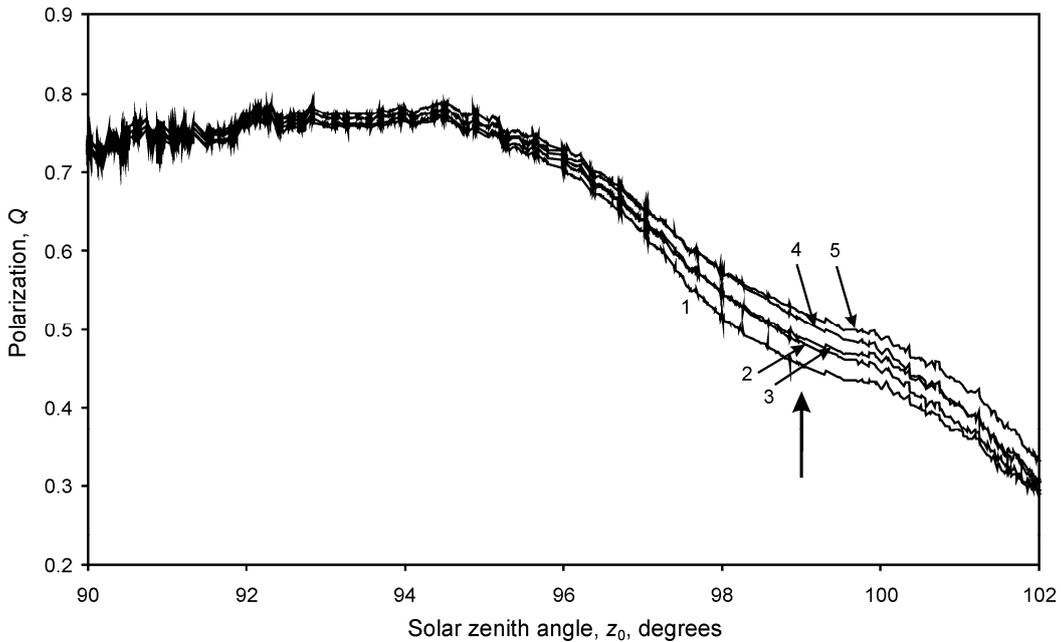

*Figure 2. Background polarization of the solar cross-vertical points ($\zeta=0$) depending on the solar zenith angle (1 – zenith ($\tau=0$), 2 – $\tau=-30°$, 3 – $\tau=+30°$, 4 – $\tau=-45°$, 5 – $\tau=+45°$). Bold arrow shows the moment of single scattering disappearance.*

The same effect in solar vertical causes the well-known sky polarization decrease, since the single scattered light is polarized horizontally there. Figure 4 shows inverted normalized second Stokes vector component $-I_2/I_1$ (in fact, the horizontal polarization) for the solar vertical points in the dusk area. Polarization drops down with $\zeta$ increase during the dark period of twilight, turning to zero at $\zeta$ about 55°. These neutral points and "reverse polarization" closer to horizon were often related with atmospheric aerosol [5], but actually are the property of multiple scattering [9, 10].

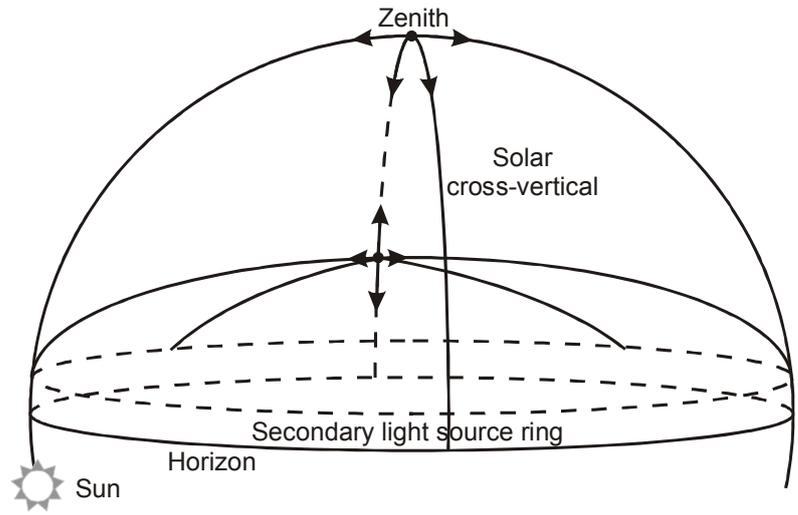

Figure 3. *"Ring" model explaining the polarization properties of multiple scattering.*

The moments of single scattering disappearance by the model [8] are shown by arrow and short vertical lines in Figure 4. They are independent on $\tau$ and defined by the empirical formula:

$$z_{0S} = 99.0 + 0.02 \cdot \zeta \qquad (1).$$

Here all values are expressed in degrees. We see that these moments correspond to the visible changes of polarization behavior in different sky points. Fast decrease of polarization as the single scattering fades away turns to its slower evolution. It is the start of "dark twilight" period [11], when all properties are defined by multiple scattering with almost constant polarization, followed by "nightfall" stage when the less polarized night sky background contribution becomes noticeable. In this paper the equation (1) will be confirmed basing on the independent study of polarization directions for the sky points away from solar vertical.

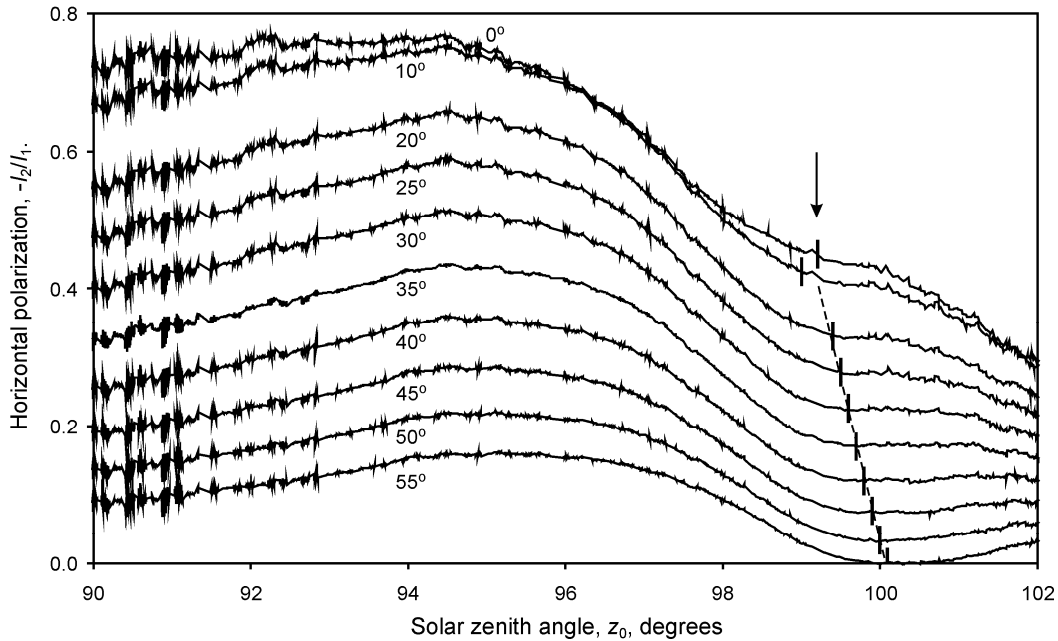

Figure 4. *Background horizontal polarization of solar vertical points in the dusk area (the values of $\zeta$ are denoted) depending on solar zenith angle. The moments of single scattering disappearance are shown.*

## 4. Polarization directions analysis and discussion.

As we saw above, "ring" model gives the brief qualitative explanation of polarization properties of multiple scattering. However, this ring is not uniform and has a strong brightness maximum in the dusk area above the Sun. It is the reason of significant polarization of multiple scattering even in the zenith, as it can be seen in Figures 2 and 4. For numerical polarization analysis we can use the model of point-like secondary light source, remembering that the position of this source can vary with the coordinates of the sky point. The multiple scattering polarization is perpendicular to the direction to this source in the sky, however, the value of polarization is less than the one for single scattering by the same angle.

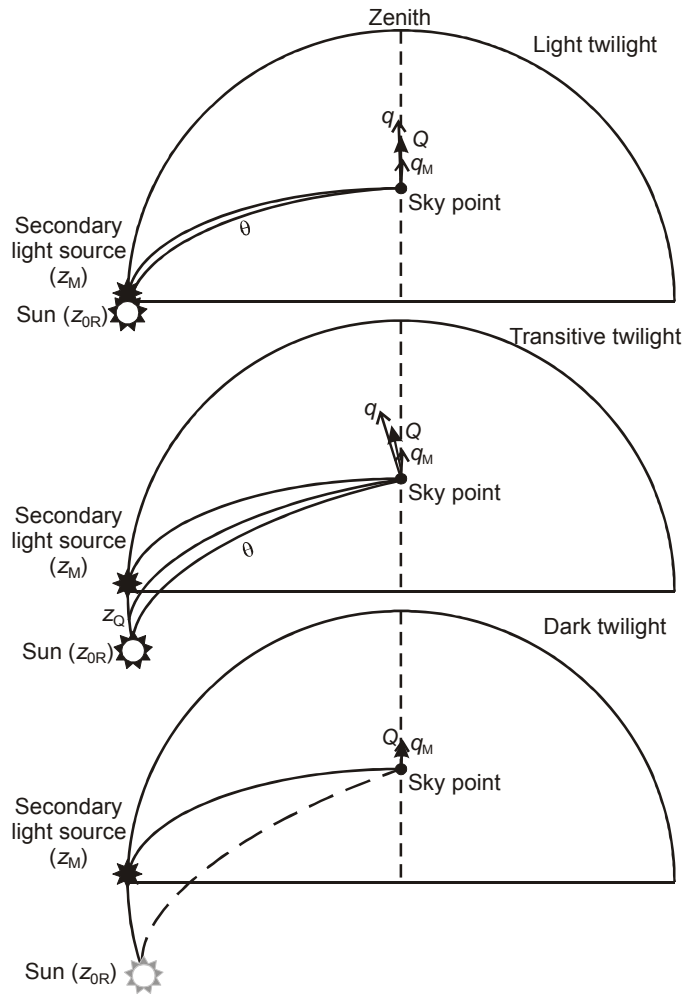

If we accept this model, than we can trace the evolution of the sky polarization direction during the twilight. We don't consider the solar vertical points, since the single and multiple scattering polarizations are collinear (or perpendicular) there. Situation is more interesting in solar cross-vertical, it is shown in Figure 5. During the light twilight, when the Sun is close to the horizon, the single scattering polarization is directed almost vertically (first graph in the figure). The direction of multiple scattering polarization is a priori unknown (as it will be shown below, it is also vertical for the solar cross-vertical points with $\zeta=0$). As the Sun depresses below the horizon, single scattering polarization direction follows it (the second graph in the figure). This deflects the total polarization direction from the vertical line. But during the transitive twilight stage single scattering disappears and the background polarization direction turns almost vertical again (third graph in the figure).

Describing it geometrically, we draw the major circle of celestial sphere across the observation point perpendicularly to the polarization direction. This circle crosses the solar vertical in the point with zenith distance $z_Q$ (shown in the second graph of Figure 5).

*Figure 5. Evolution of polarization direction in the solar cross-vertical during the twilight.*

Building the dependencies of this value on solar zenith angle, we see that real $z_Q$ behavior is the same as predicted above. Figure 6 shows it for the points with $\zeta=0°$ (cross-vertical) and $\tau=\pm45°$. Arrow shows the moment $z_{0S}=99°$ when the single scattering should disappear according to the formula (1). It is clearly confirmed here. The same is seen in the Figure 7 with the $z_Q$ dependencies for the dusk area points with $\zeta=40°$, the only difference is less value of $z_Q$ for dark twilight. The choice of threshold $z_{0S}$ value (99.8°) seems to be exact again, the same is true for other sky areas in consideration. However, we must add that these parameters can slightly change for another seasons, locations or spectral bands and must be defined by observations for any particular case.

The light source of single scattering is the Sun. Considering this component separately, we introduce the analogous zenith distance value $z_{0R}$ (Figure 5), that is equal to

$$z_{0R} = z_0 - r \qquad (2),$$

where small correction $r$ is a refraction angle for the tangent solar ray corresponding to the optical depth equal to unity (the effective twilight ray), it is about 0.1°. We denote the analogous angle $z_M$ for multiple scattering (secondary light source position). The only thing we can assume that this value is changing smoothly during the dark period of twilight.

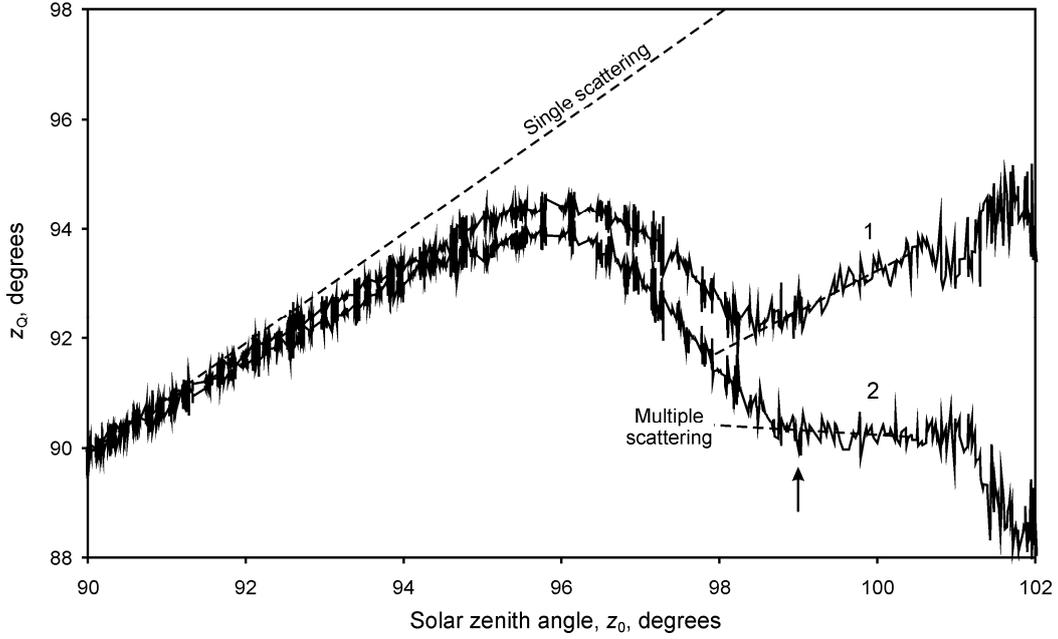

*Figure 6. Polarization direction characteristics $z_Q$ depending on the solar zenith angle for the cross-vertical points ($\zeta=0$, 1 – $\tau=-45°$, 2 – $\tau=+45°$). Arrow shows the moment of single scattering disappearance.*

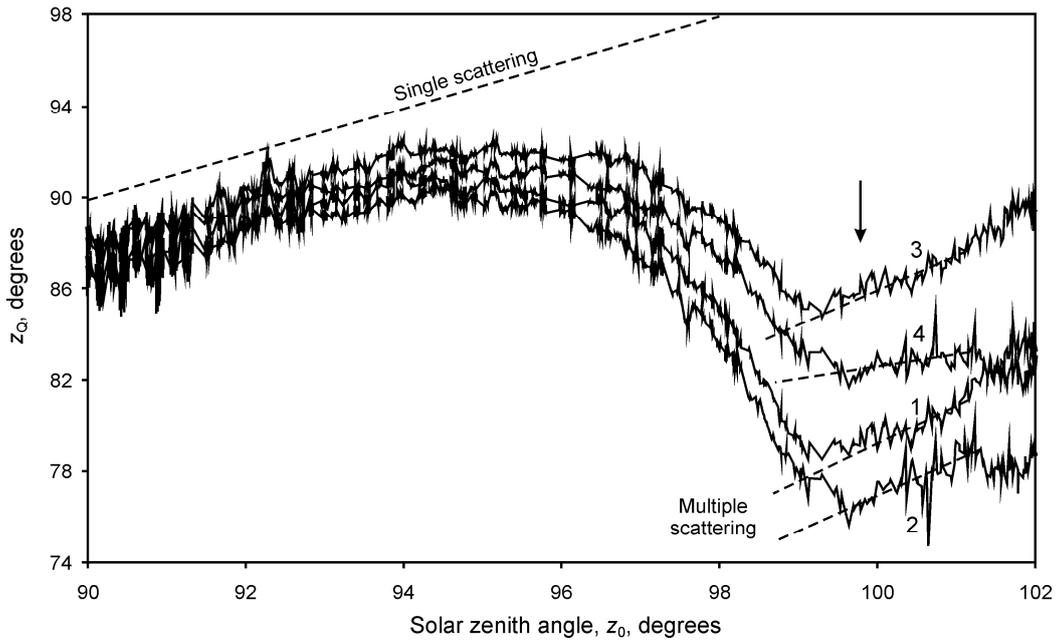

*Figure 7. Polarization direction characteristics $z_Q$ depending on the solar zenith angle for the dusk area sky points ($\zeta=+40°$, 1 – $\tau=-30°$, 2 – $\tau=+30°$, 3 – $\tau=-45°$, 4 – $\tau=+45°$). Arrow shows the moment of single scattering disappearance.*

We consider the side sky points with τ module not less than 30°. Since all $z_{Q,0R,M}$ values are close to 90° and depend linearly on the polarization angle, we can draw the simple approximate relation between them, not writing the trigonometric equations:

$$IQ = Jq + jq_M; \quad IQ \cdot z_Q = Jq \cdot z_{0R} + jq_M \cdot z_M \quad (3).$$

Here $I$ and $Q$ mean the intensity and polarization of total background, $J$ and $q$ are analogous values for single scattering, and $j$ and $q_M$ are the same for multiple scattering. Having assumed the slow evolution of $z_M$ values (linearly by solar zenith angle), and obtained them from the dark twilight period data (dashed lines in Figures 6 and 7), we can find the $Jq$ multiplication:

$$Jq = IQ \frac{z_Q - z_M}{z_{0R} - z_M} \quad (4).$$

Analogously to the intensity values $J$ in paper [8], we calculate $Jq$ for different sky points (meaning different scattering angles θ) and solar zenith angles (meaning different effective twilight layer baseline altitudes $h_B$ [8]). We can run this procedure for solar cross-vertical and dusk area, where the value of $(z_{0R} - z_M)$ is not so small. After the correction by the zenith distance cosine of the sky point, atmosphere transparency and camera flat field [8] the value obtained is the second Stokes component for a single scattering. This way we find the polarization component of single scattering matrix $(pF)_2$ depending on $h_B$. Figure 8 shows these values for the evening twilight of June, 5, 2013. Analysis by method [8] had shown the Rayleigh scattering domination (with accuracy about 1%) in the whole single scattering field for this twilight (see below). In this case $(pF)_2$ value will be proportional to pressure value $p_Q(h_B)$ on the altitude $h_B$ (in arbitrary units) and second component of Rayleigh scattering matrix $F_{2R}$:

$$(pF)_2(\theta(\zeta, \tau), h_B) = p_Q(h_B) \cdot F_{2R}(\theta) = p_Q(h_B) \sin^2\theta \quad (5).$$

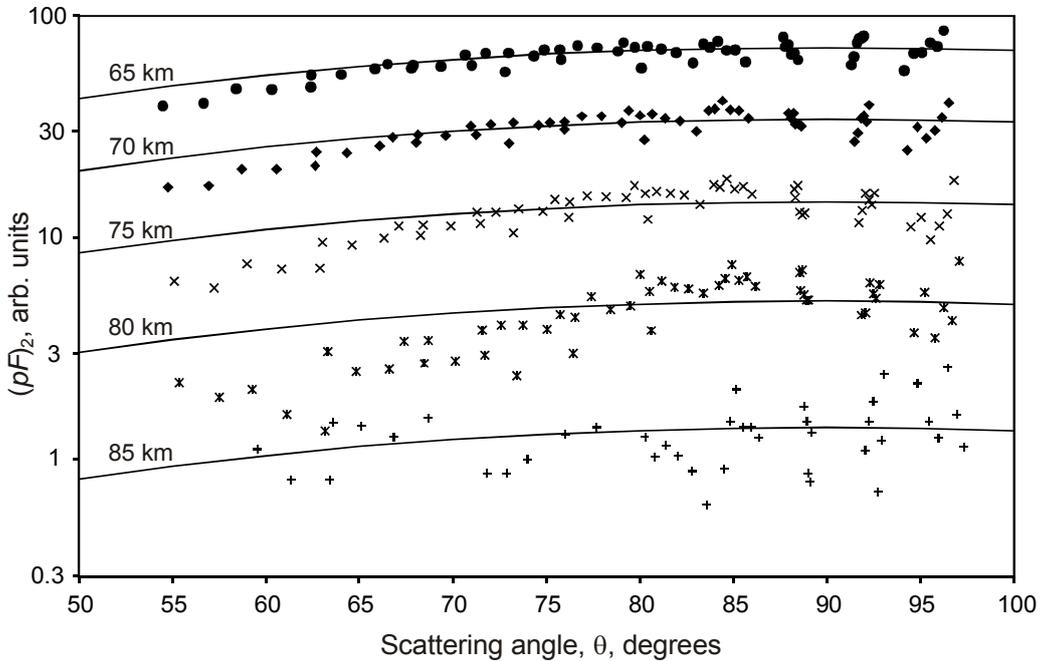

*Figure 8. Second component of Stokes vector of single scattering compared with Rayleigh curves.*

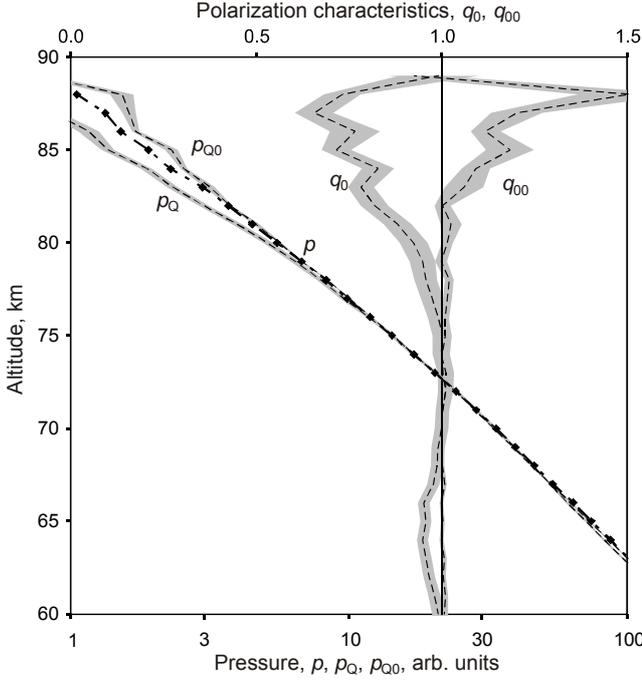
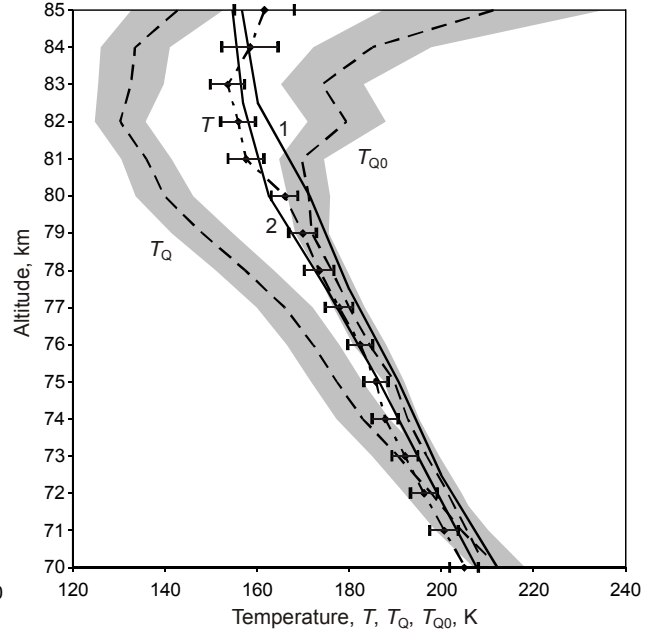

*Figure 9. Altitude profiles of pressure and polarization characteristics of single scattering.*

*Figure 10. Temperature profiles by twilight analysis compared with space data on the same date and nearby locations (1 – TIMED/SABER, 2 – EOS Aura/MLS).*

We build the angular dependencies of second component of Rayleigh scattering (solid lines in Figure 8), that are in agreement with observational data (however, with some errors). Comparing one with another, we find the pressure values $p_Q$. Taking its altitude dependencies, we can write:

$$\frac{d \ln p_Q}{dh_B} = \frac{dp_Q}{p_Q dh_B} = -\frac{\mu\, g(h_B)}{R_G T_Q(h_B)} \quad (6).$$

As in [8], here $\mu$ is the molar mass, $g$ is the gravitational acceleration, $R_G$ is the gas constant, $T_Q$ is the temperature at the altitude $h_B$. Index "Q" is used for pressure and temperature values found here in order not to mix them up with $p$ and $T$ values obtained from intensity analysis [8].

Figure 9 shows the altitude dependency of the pressure value $p$ (arbitrary units) by basic intensity method [8] and analogous values $p_{Q0}$ and $p_Q$ from polarization analysis by the methods described in [8] and this paper, respectively (for the same evening twilight of June, 5, 2013). This figure also shows the polarization characteristics values $q_{00}=p_{Q0}/p$ and $q_0=p_Q/p$ (ratios of measured and Rayleigh polarization functions of single scattering [8]). All $p$ values practically coincide for altitudes below 80 km and $q_{0,00}=1$, showing the principal Rayleigh type of single scattering. There are some errors above 80 km caused by small brightness of single scattering. These errors are more significant for $T_Q$ and $T_{Q0}$ values shown in Figure 10. In contrast, the temperature $T$ obtained by intensity analysis [8] is more exact and remarkably agrees with SABER and MLS space data for the same date and nearby location.

The fact above confirms the basic principle of twilight mesosphere study declared in [8]: temperature is better to be estimated by intensity method, which is more precise, and then polarization values can show the possible deviations of single scattering properties from the Rayleigh law and then detect the presence of depolarizing scattering substance (for example, the meteoric dust, possibly found during Perseids activity epoch [12]). The last problem can be solved now by two independent routines ([8] and this paper), this can improve the accuracy and show the principal possibility to build the common precise method based on the total amount of observational all-sky polarization data.

**5. Mesospheric temperatures and conclusion.**

Figure 11 shows the values of temperature $T$ estimated by intensity analysis [8] for WAPC observations in 2011-2014. The criteria of data selection by accuracy and Rayleigh type of single scattering are the same as in [8]. The values are compared with TIMED/SABER and EOS Aura/MLS data averaged by 8 days in time, ±3° by latitude and ±10° by longitude relatively the WAPC location, 1σ-intervals are shown. 2012 space data are selected as example. WAPC observational period covers about four months a year, but annual temperature cycle with minimum in late June – early July is noticeable.

The basic conclusion of this paper is that polarization direction, together with intensity and polarization values, provides the third information channel for separation of single and multiple scattering in the twilight sky background and study of single scattering properties in the upper atmosphere layers. Although the single scattering is two-parametric, all-sky measurements and three-dimensional polarization analysis in the points out of solar vertical can help to expand the twilight sounding possibilities.

**Acknowledgements**

Authors would like to thank T. Sergienko (Swedish Institute of Space Physics, Kiruna) for the help during the data procession.

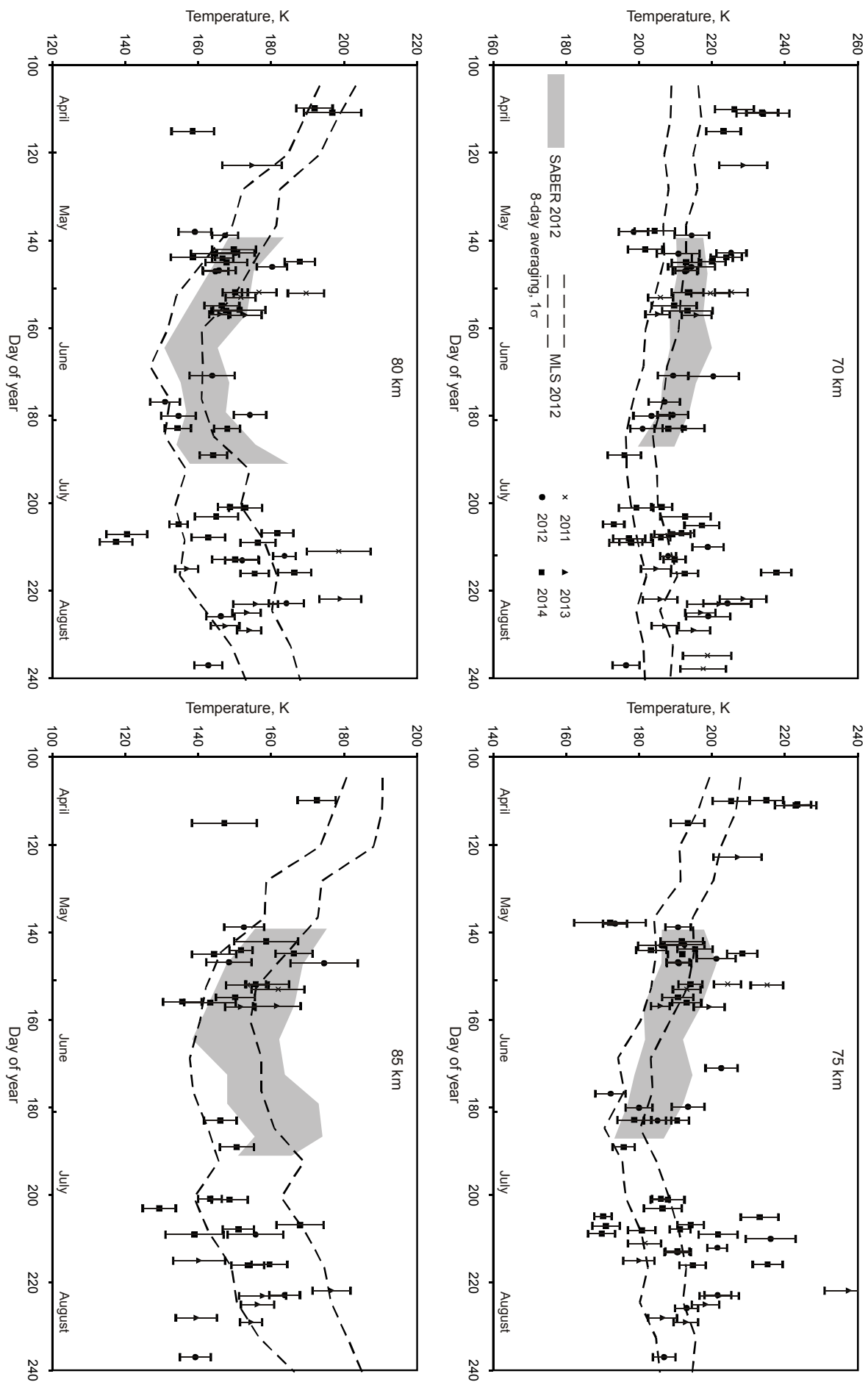

*Figure 11. WAPC temperature values in 2011-2014 compared with SABER and MLS data.*